\documentclass[aps,prl,showpacs,preprint]{revtex4}

\usepackage{graphicx} 
\usepackage{amsmath}
\usepackage{stmaryrd}
\usepackage{MnSymbol}
\usepackage{setspace}
\usepackage{physics}
\usepackage{mathtools}
\usepackage{amsfonts}
\usepackage{color}

\begin{document}

\title{Mapping of Quantum Systems to the Probability Simplex}
\author{D. D. Yavuz and A. Yadav}
\affiliation{Department of Physics, 1150 University Avenue,
University of Wisconsin at Madison, Madison, WI, 53706}
\date{\today}
\begin{abstract}

We start with the simplest quantum system (a two-level system, i.e., a qubit) and discuss a one-to-one mapping of the quantum state in a two-dimensional Hilbert space to a vector in an eight dimensional probability space (probability simplex). We then show how the usual transformations of the quantum state, specifically the Hadamard gate and the single-qubit phase gate, can be accomplished with appropriate transformations of the mapped vector in the probability simplex. One key defining feature of both the mapping to the simplex and the transformations in the simplex is that they are not linear. These results show that both the initial state and the time evolution of a qubit can be fully captured in an eight dimensional probability simplex (or equivalently using three classical probabilistic bits). We then discuss multi-partite quantum systems and their mapping to the probability simplex. Here, the key tool is the identical tensor product structure of combining multiple quantum systems as well as multiple probability spaces. Specifically, we explicitly show how to implement an analog of the two-qubit controlled-not (CNOT) gate in the simplex. We leave it an open problem how much the quantum dynamics of $N$ qubits can be captured in a probability simplex with $3N$ classical probabilistic bits. Finally, we also discuss the equivalent of the Schrodinger's equation for the wavefunction (in a Hilbert space of arbitrary dimension), which dictates the time evolution of the vectors in the simplex. 

\end{abstract} 
\maketitle

\section{I. Introduction}

The birth of quantum mechanics has fundamentally changed our understanding and interpretation of the natural laws of the microscopic world. The mathematical structure and the experimental predictions of the theory have been quite clear; yet, there has been a wide-ranging and still ongoing rigorous debate regarding the precise meaning and interpretation of various key components of the theory, such as the wavefunction and the measurement \cite{bell1,fuchs1}.  According to the traditional Copenhagen interpretation, the evolution of the quantum wavefunction is described by the Schrodinger's equation, which is deterministic. However, when a measurement is performed, the deterministic evolution comes to an abrupt end with probabilities entering the picture. The experimenter observes probabilistic outcomes, with the likelihood of each outcome determined by the overlap of the wavefunction (at the time of the measurement) with the eigenstates of the basis that the measurement is performed at. 

This probabilistic nature of the quantum theory is now very well understood, and has been confirmed by countless experiments. Yet, the mathematical structure of quantum mechanics is quite different compared to a traditional theory of probability. This has been emphasized in detail by the recent pioneering work of Fuchs and colleagues \cite{fuchs1}. Specifically, quantum mechanics invokes the mathematical structure of wavefunctions with complex coefficients that live in a Hilbert space. A key question is whether this unique mathematical structure is strictly necessary. Can quantum mechanics be interpreted under the general umbrella of classical probabilistic theories? This question is not only important from a foundational point of view, but also has practical implications. Recent achievements in quantum computing and quantum information science have shown us that quantum devices have the potential to make a large impact on society \cite{nielsen,divincenzo1,divincenzo2,vazirani,shor,ekert}. However, to be able to understand the true power of quantum devices, we need to better understand the boundary between quantum mechanics, classical physics, and traditional probabilistic theories \cite{tsang,watrous,wetterich1,wetterich2}. To answer this great need, over the last two decades, there has been an impressive body of research that has investigated the relationship between quantum mechanics and classical probabilistic theories. One line of research has argued for a completely subjective interpretation of the wavefunction \cite{fuchs1,fuchs2}. One of the main tools in these investigations is fine-tuned operator classes that allow Symmetric Informationally Complete (SIC) measurements \cite{fuchs3,boyd}. Other related research has tried to place quantum mechanics under the umbrella of probability theories that are more general than classical, sometimes referred to as post-classical theories of probability \cite{hardy,barrett1,barnum,masanes,rau}. This research has identified a rich landscape and the goal is to place quantum mechanics properly in this landscape in order to better understand its unique properties. 

The main goal of this paper is to point out another direction in investigating the relationship between quantum mechanics and classical probabilistic theories: maps from the Hilbert space to a probability space and transformations in the same space, with the unique feature that the maps and transformations are not necessarily linear. In this work, we focus on capturing the initial state and the time evolution of quantum systems in a classical probability space (i.e., we will not consider the measurement aspect of quantum systems \cite{realq1,realq2}).  Specifically, we start with the simplest quantum system (a two-level system, i.e., a qubit) and discuss a mapping of the quantum state to a vector in a probability space (Fig.~1). The mapping is one-to-one and preserves all the information encoded in the wavefunction. Not surprisingly, to be able to store all the information encoded in the complex coefficients, we need to increase the dimension of the system: the mapping is to an eight dimensional probabilistic space (from the two dimensional Hilbert space). 

Once a quantum state is mapped, the next key question is whether the evolution of the state can be captured in the probability simplex. It is well known that an arbitrary evolution of a single qubit can be achieved using combinations of Hadamard gates and phase rotations \cite{nielsen}. We will show how these two main operations can be implemented with appropriate transformations of the mapped vector in the probability space. The transformations in the simplex are affine, but not linear. 

These results show that both the initial state and the time evolution of a single qubit can be captured in an eight dimensional probability space (or equivalently using three classical probabilistic bits). We then investigate this mapping for multiple qubits. Here, the key tool is the identical tensor product structure of combining multiple quantum systems and combining multiple probability spaces. Specifically, we will explicitly discuss how to implement an analog of the two-qubit controlled-not (CNOT) gate in the simplex. Within our formulation,  $N$ qubits will map to a tensor product of $N$ 8-dimensional probability spaces (i.e., a Hilbert space of dimension $2^N$ will map to a probability simplex of dimension $8^N$). We leave it as an open question how much of the dynamics of $N$ qubits can be captured in the simplex with $3N$ classical probabilistic bits. Finally, we will also discuss an analogue of the Schrodinger's equation for the wavefunction which lives in a Hilbert space of arbitrary dimension. This is a continuous differential equation that describes the evolution of the simplex vector under an effective ``Hamiltonian".  

Our work has been heavily influenced by the recent investigations of quantum mechanics within the operational framework of probability theories; in particular the pioneering work of Hardy \cite{hardy} and Barrett \cite{barrett1}. There is also a relation of this work to a large body of literature who have attempted to derive some features of quantum mechanics using classical ``toy" theories. A good summary of various toy theories is discussed in, for example, Ref.~\cite{rudolph1}.   This work is also related to the mapping of quantum states to probability-like distributions, typically referred to as quasiprobabilities \cite{bartlett1,bartlett2,bartlett3,raussendorf1,raussendorf2,eisert,zhu,wootters}. The most well-known example of a quasiprobability distribution is the Wigner function. It is well-known that quasi-probabilities can have negative values; in fact, the true quantum mechanical nature of the wavefunction is expressed in these negative regions. We argue that when one allows for maps and transformations that are not necessarily linear, one can capture a quantum state (as well as its' evolution) using only probabilities (i.e., negative values are not needed). We will comment on these connections more thoroughly below, in Section~XIV of the paper. 

\begin{figure}[tbh]
\vspace{-0cm}
\begin{center}
\includegraphics[width=13cm]{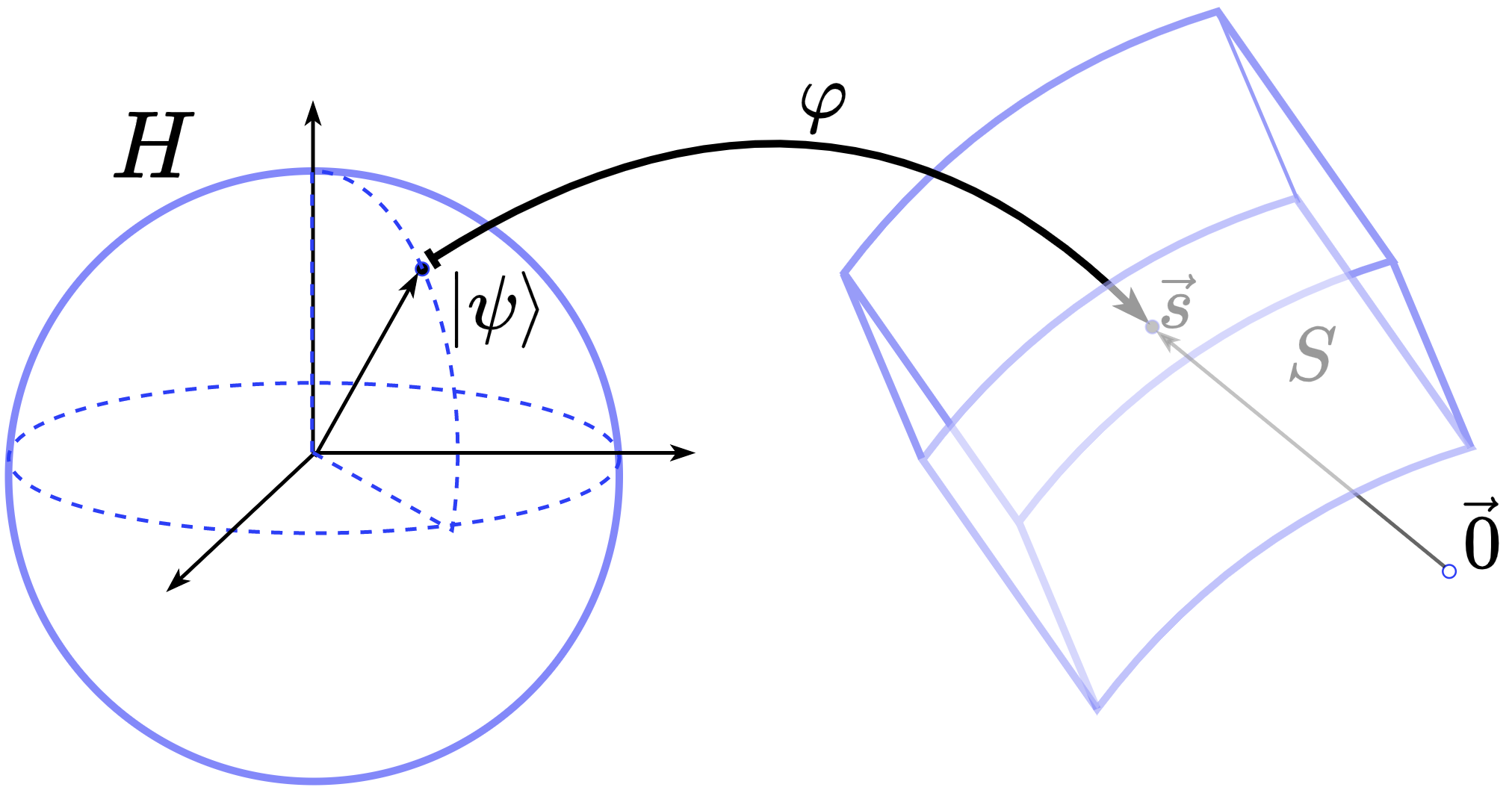}
\vspace{-0.2cm} 
\caption{\label{scheme} \small The schematic of the basic idea that we will study in this work. We start with a qubit with wavefunction, $| \psi \rangle $, and discuss a one-to-one mapping, $\varphi$ from the Hilbert space $H$ to a vector $\vec{s}$ in an eight dimensional probability space $S$ (which is a real Euclidean space). The vectors $\vec{s}$ have their values add up to unity (i.e., they are valid probability distributions), and therefore lie on a hypersurface (the simplex). We then show how various elementary gates that are applied to the wavefunction, $| \psi \rangle$, can be mimicked in the probability simplex (i.e., not just the initial quantum state, but also its evolution can be mapped to the simplex). }
\end{center}
\vspace{-0.3cm}
\end{figure}

We note that there is nothing new in the idea that a quantum state can be represented as a vector of probabilities. Consider a specific set-of basis states, $|\phi_1 \rangle$, $|\phi_2 \rangle$, $|\phi_3 \rangle$, ..... Given an initial wavefunction $|\psi \rangle$, we can view the measurement outcomes in the mentioned basis as a probabilistic vector of the form, $ \left( | \langle \phi_1 | \psi \rangle|^2, | \langle \phi_2 | \psi \rangle|^2, | \langle \phi_3 | \psi \rangle|^2, ... \right) $. When the quantum state evolves, these probability values would change, and we can always view this as mapping the initial probability vector to another output probability vector (by multiplying with an appropriate matrix, for example). Such straightforward mapping, however, always results in matrices that {\it depend} on the initial wavefunction $|\psi \rangle$. This is, for example, clearly discussed in the ontological model of Aaronson, in Ref.~\cite{aaronson}

What we will be discussing is something quite different. We will encode both the amplitude and phase information of the complex wavefunction in probabilities. The analogs of the quantum gates in the probability space (for example the Hadamard gate and the single-qubit phase gate), will involve affine transformations, with constant translations and constant matrices (that will be {\it independent} of the initial input state). This essentially means that all the interferences in the complex coefficients can be captured in the probability space. To further emphasize the analogy, consider a single qubit that evolves through any number of gates: for example, a Hadamard gate, followed by a phase-gate, followed by $\pi/3$ Rabi rotation, and so forth. One can construct circuits that would perform exact analogs of these gates in the probability simplex, and exactly track the quantum evolution in the simplex. Once the circuits are fixed, if the initial condition for the quantum state changes, one only needs to modify the initial condition for the simplex vector.

\section{II. Preliminaries}
As we mentioned above, in traditional formulation of quantum mechanics, a quantum state is described by a complex wavefunction, $|\psi \rangle$, in a Hilbert space, $H$. This state will evolve according to Schrodinger's equation, which conserves the norm of the wavefunction. This time evolution of the quantum state can be described using an appropriate unitary matrix, $\hat{U}$, that satisfies, $\hat{U}^\dagger \hat{U} = \hat{U} \hat{U}^\dagger = \hat{I}$. With this evolution, the state is mapped to $|\psi \rangle \longrightarrow \hat{U} | \psi \rangle$. 

In any classical probabilistic experiment, we will have a set of probabilities, which we can also think of as constituting a vector, in a probabilistic space, $S$. We will denote such a probabilistic vector with $\vec{s}$. Each of the entries of this vector has to be between 0 and 1, i.e., $0 < s_i <1$, and furthermore, the entries need to sum to unity, $\sum_i s_i =1$. Because the entries add up to unity, such a vector lies on certain surface in the probabilistic space, and this surface is called the simplex \cite{definetti}. Similar to a quantum state, such a probabilistic vector can also evolve in time (perhaps the experimental conditions change, or our knowledge of the experiment evolves). We can view such evolution as mapping a vector in space $S$, to another vector. We will denote such mapping with $ T : S \longrightarrow S$. Usually, such evolution is described by multiplying the vector $\vec{s}$ with a Stochastic matrix, $\tilde{\mathcal{M}}$, i.e., $T (\vec{s}) = \tilde{\mathcal{M}} \cdot \vec{s}$. A stochastic matrix is a matrix whose columns sum up to 1. This assures that the resultant vector also is normalized; i.e., its' components add up to unity. As we will discuss below in detail, probabilistic vectors can also transform using maps that are not linear, but instead affine. Throughout this paper, all quantum mechanical operators will be presented by a hat (for example $\hat{U}$), whereas all the  transformations in the simplex will be presented by a tilde (for example, $\tilde{\mathcal{M}}$).

\section{III. Two-level quantum system (qubit) and mapping}

We now specifically focus on a single qubit wavefunction $ | \psi \rangle = c_0 | 0 \rangle + c_1 | 1 \rangle $. Here, the states $| 0 \rangle $ and $| 1 \rangle$ are the logical states, and  $c_0$ and $c_1$ are the complex coefficients satisfying the usual normalization condition, $| c_0|^2 + |c_1|^2 =1$. The quantum state can be represented as a vector in the logical state basis:
\begin{eqnarray}
| \psi \rangle = 
\left( \begin{array}{c} 
c_0 \\
c_1 
\end{array} \right) \equiv 
\left( \begin{array}{c} 
x_0 + i y_0 \\
x_1 + i y_1
\end{array} \right) \quad . 
\end{eqnarray}

\noindent Here, we have defined the real and imaginary parts of each complex coefficient, i.e., $c_0 \equiv x_0 + i y_0$ and $c_1 \equiv x_1 + i y_1$.  In what follows, instead of the complex coefficients, we will work with these real and imaginary parts. This is necessary since the probabilities are real and positive, and therefore, components of a vector in the probability simplex must only contain real coefficients.

We propose the following mapping of the quantum state $|\psi \rangle$ (in Hilbert space $H$) to a vector $\vec{s}$ in the probability space, $\varphi:H\rightarrow S$ :
\begin{eqnarray}
| \psi \rangle = 
\left( \begin{array}{c} 
x_0 + i y_0 \\
x_1 + i y_1
\end{array} \right) \xrightarrow{\varphi}
\vec{s}= \frac{1}{8} \left( \begin{array}{c} 
1+x_0 \\
1+x_1 \\
1-x_0 \\
1-x_1 \\
1+y_0 \\
1+y_1 \\
1-y_0 \\
1-y_1
\end{array} \right)  \quad .
\end{eqnarray}

We note that the vector $ \vec{s}$ represents a valid probability distribution. That is, each of the entries is between 0 and 1 (i. e., $ 0 < s_i <1$), and these entries sum up to unity, $\sum_i s_i =1$. The fact that we need to increase the dimension from 2 to 8 is intuitive. For each complex coefficient, we need to store two real numbers, the real part and the imaginary part. Furthermore, for each real number, we need to store the quantity with both signs. This is because, in order to map the transformations of the quantum state, we will need access to both signs of these coefficients. Hence, the factor of 4 increase in the dimension. The map is injective (i.e., one-to-one), but not surjective. The main insight in the mapping of Eq.~(2) is that the phase and the amplitude information (for the real and imaginary parts of the complex coefficients) can be stored in how much the probabilities deviate from purely random quantity (hence the initial ``1" in all the entries of $\vec{s}$). 

A key property of the mapping of Eq.~(2) is that it is not linear. By inspection, a superposition of two wavefunctions do not map to the same superposition of their mapped vectors: $\varphi(a\ket{\psi}+b\ket{\phi})\neq a\varphi(\ket{\psi})+b\varphi(\ket{\phi})$ for $\ket{\psi},\ket{\phi}\in H; a,b\in \mathbb{C}$.

We note that the set of states $\vec{s}$ defined in the simplex by Eq.~(2) form a convex surface. That is, for two different states $\vec{s}$ and $\vec{s}'$, and for coefficients $\lambda$ and $\lambda'$ such that $\lambda + \lambda' =1$, any combination $ \lambda \vec{s} + \lambda' \vec{s}'$ is also an allowed mapped state. This is similar to what is discussed in Refs.~\cite{barrett1,hardy}. We also note, however, that, differing from the prior work, the simplex vector with all of its' entries equal to 0 (which we can denote by $\vec{0}$) is not a valid mapped vector. Even if we were to include not-normalized quantum states (where the probabilities leek out of the system, for example), in the limit, $x_i \rightarrow 0, y_i \rightarrow 0$, all of the entries for the vector in the simplex would approach $\frac{1}{8}$, i.e., $ \vec{s} \rightarrow \frac{1}{8} \vec{1}$. 

\section{iV. Transformations in the simplex}

As we mentioned above, prior work has focused on linear transformations of probability vectors of the form $\tilde{\mathcal{M}} \cdot \vec{s}$, using a stochastic matrix $\tilde{\mathcal{M}}$.  There is an important reason for the consideration of only linear transformations in earlier work (for example, in Refs.~\cite{hardy} and \cite{barrett1}. From the construction of the probability space,  these authors included the null vector, $\vec{0}$, in their allowed states and further posed the constraint that a reasonable map, $f$, should leave the null vector unchanged, i.e., $f(\vec{0}) \rightarrow \vec{0}$. However, as we discussed above, the null vector,  $\vec{0}$,  in the probability simplex is not a valid mapped vector in our formalism. 

Furthermore, operations of probabilities that are not linear are encountered frequently and we do not see an apriori reason to exclude such maps. One well-known example is the nonlinearity due to the logical OR operation. Consider two independent events, $A$ and $B$, with event probabilities, $p(A)$ and $p(B)$, respectively. The probability of $A$ OR $B$ would be given by, $p(A \lor B) = p(A) + p(B) - p(A \land B) = p(A) + p(B) - p(A) p(B)$, which is a nonlinear function in each of the event probabilities. 

The central question is what type of maps, $T : S \rightarrow S$, that we should be looking for. Motivated by the mapping of Eq.~(2), we look for affine transformations of the simplex vector of the form a translation added on linear combinations of the simplex vector entries.  A key insight of this work is that the mapped vectors [described by Eq.(2)] can be written in the following form, which makes the correspondence with the Hilbert space more clear. 

\begin{eqnarray}
\vec{s} & = & \frac{1}{8} \left( \begin{array}{c} 
1 \\
1 \\
1 \\
1 \\
1 \\
1 \\
1 \\
1 
\end{array} \right) +
\frac{1}{8} \left( \begin{array}{r} 
x_0 \\
x_1 \\
-x_0 \\
-x_1 \\
y_0 \\
y_1 \\
-y_0 \\
-y_1 
\end{array} \right)  \nonumber \\
\Rightarrow \vec{s} & \equiv &  \frac{1} {8} \left( \vec{1} + \vec{p} \right) \quad . 
\end{eqnarray}
Here, we have defined another 8-dimensional vector, $\vec{p} \equiv 8 \vec{s}-\vec{1}$. Note that the entries of $\vec{p}$ sum up to zero; i.e., $\sum_i p_i =0$. Furthermore, the Euclidian norm of $\vec{p}$ is a constant $||\vec{p}||=\sqrt{2}$, since we have $x_0^2+y_0^2+x_1^2+y_1^2 =1$ (this is because of the normalization of the wavefunction). We also note that the two vectors that form the simplex vector $\vec{s}$ are orthogonal to each other, $\vec{1} \cdot \vec{p} =0$. As a result, we have $ ||\vec{s}||=\sqrt{||\vec{1}||^2+||\vec{p}||^2}/8=\sqrt{10}/8$, which is also constant. This shows that $\vec{s}$ lies on the intersection of a seven-dimensional hypersphere, with four seven-dimensional hyperplanes, resulting in a three dimensional hypersurface $S$. 

As it will be clear below, because the quantum gates form linear combinations of the entries of $\vec{p}$, we first view the mapping of the simplex vector $\vec{s}$, as instead mapping $\vec{p}$ to another vector. We will call the matrix for this mapping to be $\tilde{M}$:
\begin{eqnarray}
\vec{p} \longrightarrow \tilde{M} \cdot \vec{p} \quad . 
\end{eqnarray}
Expressed as acting on the full simplex state $\vec{s}$, the transformation of Eq.~(4) is, $T(\vec{s}) = \frac{1}{8} \left( \vec{1} + \tilde{M} \cdot \vec{p} \right)$, which gives $ T(\vec{s}) = \frac{1}{8} \left[ \vec{1} + \tilde{M} \cdot \left( 8 \vec{s} -  \vec{1} \right) \right] $, or writing it slightly differently, 
\begin{eqnarray}
T(\vec{s}) =\frac{1}{8} \left( \tilde{I} - \tilde{M} \right) \cdot \vec{1} + \tilde{M} \cdot \vec{s} \quad .  
\end{eqnarray}
\noindent Here, the quantity $\tilde{I}$ is $ 8 \times 8$ identity matrix. Below, we will give explicit expressions for the $8 \times 8$ matrices, $\tilde{M}$, for specific evolution of the quantum state, such as the application of the Hadamard gate. With the matrix $\tilde{M}$ given, Eq.~(5) describes the explicit transformation of the probability vector, with the map $T : S \longrightarrow S$ in the simplex. 

We note that, the first term in the right hand side of Eq.~(5) is a translation for each of the entries of the vector (an offset). Because of this term, the map $T : S \longrightarrow S$ is not linear (i.e., the sum of two vectors $\vec{s}$ and $\vec{s'}$ would not transform as the sum of the individual transforms). However, $T$ is an affine map. For two vectors, $\vec{s}$ and $\vec{s}'$, and for coefficients $\lambda$ and $\lambda'$ such that $\lambda + \lambda' =1$, we have $ T( \lambda \vec{s} + \lambda' \vec{s}') = \lambda T(\vec{s}) + \lambda' T(\vec{s}') $. 

The constraints on the matrix $\tilde{M}$ of above such that $T : S \longrightarrow S$  is a valid map is different from stochasticity. Specifically, the two necessary constraints are (1) $\tilde{M}$ should be such that the norm of the resulting vector is preserved since we need to have: $ || \tilde{M} \cdot \vec{p} || = \sqrt{2}$. Because of the specific form for the vector $\vec{p}$, this norm conservation does not imply orthogonality of the matrix $\tilde{M}$. By inspection, the necessary constraint is that the sum of the squares of the entries in each row must add up to unity: i.e., $\sum_j \tilde{M}_{{ij}}^2 =1$ for each row $i$.  (2) The rows of $\tilde{M}$ should be related to each other such that the entries of $ \tilde{M} \cdot \vec{p}$ sum up to zero. Specifically, $ \tilde{M} \cdot \vec{p}$ should produce a column vector of the form shown in Eq.~(2), with respective entries having equal amplitude and opposite signs.   This assures that the resulting full simplex vector, $\frac{1}{8} \left( \vec{1} +\tilde{M} \cdot \vec{p} \right)$ is  a valid probability distribution (i.e., its' entries add up to unity).

We next discuss how using the above affine transformations, one can find the equivalents of the Hadamard gate, Rabi rotations, and the single-qubit phase gate in the probabilistic simplex. 

\section{V. Hadamard gate}

The Quantum Mechanical Hadamard gate on a single qubit is accomplished by multiplying the state vector $| \psi \rangle$ with the following unitary matrix \cite{nielsen}:
\begin{eqnarray}
\hat{U}_H = \left(\begin{array}{rr} 
\frac{1}{\sqrt{2}} & -\frac{1}{\sqrt{2}} \\
\frac{1}{\sqrt{2}} & \frac{1}{\sqrt{2}} 
\end{array} \right) \quad .
\end{eqnarray}

\noindent Note that as required by unitarity, we have $\hat{U}_H^\dag \hat{U}_H=1$. The effect of the Hadamard gate on the quantum state, explicitly expressed in terms of the real and imaginary parts of the complex coefficients, is:
\begin{eqnarray}
\hat{U}_H | \psi \rangle & = &  \left( \begin{array}{rr} 
\frac{1}{\sqrt{2}} & -\frac{1}{\sqrt{2}} \\
\frac{1}{\sqrt{2}} & \frac{1}{\sqrt{2}} 
\end{array} \right)  
\left( \begin{array}{c} 
x_0 + i y_0 \\
x_1 + i y_1
\end{array} \right) \nonumber \\
&  =  & \left( \begin{array}{c} 
\frac{1}{\sqrt{2}} x_0 -   \frac{1}{\sqrt{2}} x_1 + i \left( \frac{1}{\sqrt{2}} y_0 -  \frac{1}{\sqrt{2}} y_1 \right) \\
\frac{1}{\sqrt{2}} x_0 +  \frac{1}{\sqrt{2}} x_1 + i \left( \frac{1}{\sqrt{2}} y_0 +  \frac{1}{\sqrt{2}} y_1 \right) 
\end{array} \right)  \quad . 
\end{eqnarray}

\noindent We, therefore, have the following transformation of the quantum state as a result of the Hadamard gate:
\begin{eqnarray}
| \psi \rangle & \longrightarrow & \hat{U}_H | \psi \rangle \nonumber \\
\left( \begin{array}{c} 
x_0 + i y_0 \\
x_1 + i y_1
\end{array} \right) & \longrightarrow &  
\left( \begin{array}{c} 
\frac{1}{\sqrt{2}} x_0 -   \frac{1}{\sqrt{2}} x_1 + i \left( \frac{1}{\sqrt{2}} y_0 -  \frac{1}{\sqrt{2}} y_1 \right) \\
\frac{1}{\sqrt{2}} x_0 +  \frac{1}{\sqrt{2}} x_1 + i \left( \frac{1}{\sqrt{2}} y_0 +  \frac{1}{\sqrt{2}} y_1 \right) 
\end{array} \right)  \quad . 
\end{eqnarray}

\noindent The central question is if a similar transformation can be accomplished in the simplex, using an appropriate matrix, which we will denote with $\tilde{M}(\hat{U}_H)$. Specifically, we are looking for the following transformation:
\begin{eqnarray}
\vec{p} & \longrightarrow &  \tilde{M}(\hat{U}_H) \cdot \vec{p} \nonumber \\
 \left( \begin{array}{r} 
x_0 \\
x_1 \\
-x_0 \\
-x_1 \\
y_0 \\
y_1 \\
-y_0 \\
-y_1 
\end{array} \right) & \longrightarrow & 
\left( \begin{array}{r} 
\left( \frac{1}{\sqrt{2}} x_0 -   \frac{1}{\sqrt{2}} x_1 \right)  \\
\left( \frac{1}{\sqrt{2}} x_0 +  \frac{1}{\sqrt{2}} x_1 \right)  \\
- \left( \frac{1}{\sqrt{2}} x_0 -   \frac{1}{\sqrt{2}} x_1 \right) \\
-\left( \frac{1}{\sqrt{2}} x_0 +  \frac{1}{\sqrt{2}} x_1 \right)  \\
\left( \frac{1}{\sqrt{2}} y_0 -   \frac{1}{\sqrt{2}} y_1 \right)\\
\left( \frac{1}{\sqrt{2}} y_0 +  \frac{1}{\sqrt{2}} y_1 \right) \\
- \left( \frac{1}{\sqrt{2}} y_0 -   \frac{1}{\sqrt{2}} y_1 \right)\\
- \left( \frac{1}{\sqrt{2}} y_0 +   \frac{1}{\sqrt{2}} y_1 \right)
\end{array} \right) \quad . 
\end{eqnarray}

\noindent By inspection, the required $ 8 \times 8$ matrix for the transformation of Eq.~(9) is:
\begin{eqnarray}
\tilde{M}(\hat{U}_H) = \left( \begin{array}{cccccccc}
\frac{1}{\sqrt{2}} & 0 & 0 & \frac{1}{\sqrt{2}} &  0 & 0 & 0 & 0 \\
\frac{1}{\sqrt{2}} & \frac{1}{\sqrt{2}} & 0 & 0 & 0 & 0 & 0 & 0 \\ 
0 & \frac{1}{\sqrt{2}} & \frac{1}{\sqrt{2}} & 0 & 0 & 0 & 0 & 0 \\
0 & 0 & \frac{1}{\sqrt{2}} & \frac{1}{\sqrt{2}} & 0 & 0 & 0 & 0 \\
0 & 0 & 0 & 0 & \frac{1}{\sqrt{2}} & 0 & 0 & \frac{1}{\sqrt{2}} \\
0 & 0 & 0 & 0 & \frac{1}{\sqrt{2}} & \frac{1}{\sqrt{2}} & 0 & 0 \\
0 & 0 & 0 & 0 & 0 & \frac{1}{\sqrt{2}} & \frac{1}{\sqrt{2}} & 0 \\
0 & 0 & 0 & 0 & 0 & 0 & \frac{1}{\sqrt{2}} & \frac{1}{\sqrt{2}} \\
\end{array} \right) \quad . 
\end{eqnarray}

With the matrix $\tilde{M}(\hat{U}_H)$ given as above, the full transformation of the simplex vector is given by Eq.~(5), i.e., $T_{\hat{U}_{H}}(\vec{s}) =  \frac{1}{8} \left[ \tilde{I} - \tilde{M}(\hat{U}_H) \right] \cdot \vec{1} + \tilde{M}(\hat{U}_H) \cdot \vec{s}$. 

\section{VI. Rabi rotations}

Above in Section~V, we discussed the Hadamard gate. However, any arbitrary qubit rotation in that plane can be accomplished in a similar way: i.e., one can observe ``Rabi flopping" of a single qubit by implementing the following rotation in the simplex:
\begin{eqnarray}
\hat{U}_\theta = \left( \begin{array}{rr} 
\cos{\theta} & -\sin{\theta} \\
\sin{\theta} & \cos{\theta}
\end{array} \right) \quad .
\end{eqnarray}

\noindent Here, the quantity $\theta$ is the Rabi rotation angle. For $\theta = \pi/4$, we recover the Hadamard gate of above. The effect of this rotation on the quantum state is:
\begin{eqnarray}
\hat{U}_\theta | \psi \rangle & = &  \left( \begin{array}{rr} 
\cos{\theta} & -\sin{\theta} \\
\sin{\theta} & \cos{\theta}
\end{array} \right)  
\left( \begin{array}{c} 
x_0 + i y_0 \\
x_1 + i y_1
\end{array} \right) \nonumber \\
&  =  & \left( \begin{array}{c} 
x_0 \cos{\theta}  -   x_1 \sin{\theta}  + i \left( y_0 \cos{\theta} -  y_1 \sin{\theta}   \right) \\
x_0 \sin{\theta} +   x_1 \cos{\theta} + i \left( y_0 \sin{\theta}  +  y_1 \cos{\theta}   \right) 
\end{array} \right)  \quad . 
\end{eqnarray}

\noindent The question again is if a similar transformation can be accomplished in the simplex, using an appropriate matrix, which we term $\tilde{M} (\hat{U}_\theta )$.  By inspection, the required $ 8 \times 8$ matrix for this transformation is:
\begin{eqnarray}
\tilde{M} (\hat{U}_\theta ) = \left( \begin{array}{cccccccc}
\cos{\theta} & 0 & 0 & \sin{\theta} &  0 & 0 & 0 & 0 \\
\sin{\theta} & \cos{\theta} & 0 & 0 & 0 & 0 & 0 & 0 \\ 
0 & \sin{\theta} & \cos{\theta} & 0 & 0 & 0 & 0 & 0 \\
0 & 0 & \sin{\theta} & \cos{\theta} & 0 & 0 & 0 & 0 \\
0 & 0 & 0 & 0 & \cos{\theta} & 0 & 0 & \sin{\theta} \\
0 & 0 & 0 & 0 & \sin{\theta} & \cos{\theta} & 0 & 0 \\
0 & 0 & 0 & 0 & 0 & \sin{\theta} & \cos{\theta} & 0 \\
0 & 0 & 0 & 0 & 0 & 0 & \sin{\theta} & \cos{\theta} \\
\end{array} \right) \quad . 
\end{eqnarray}

With this transformation matrix, the individual entries of the simplex vector would oscillate as a function of the rotation angle $\theta$. An example is shown in Fig.~2. Here we perform the transformation on the simplex vector,  $ T_{\hat{U}_{\theta}}(\vec{s}) = \frac{1}{8} \left[ \tilde{I} - \tilde{M} (\hat{U}_\theta ) \right] \cdot \vec{1} + \tilde{M} (\hat{U}_\theta ) \cdot \vec{s}$, with the Rabi rotation matrix $\tilde{M} (\hat{U}_\theta )$ of Eq.~(13), with the initial condition of $x_0=1, y_0=0, x_1=0, y_1=0$ for the quantum wavefunction (and therefore the simplex amplitudes). As the rotation angle $\theta$ is varied, the first probability component of the simplex vector $s_1= \frac{1}{8} (1+x_0 \cos{\theta} - x_1 \sin{\theta})$ oscillates between the values of $0$ and $\frac{1}{4}$.  

\begin{figure}[tbh]
\vspace{-0cm}
\begin{center}
\includegraphics[width=15cm]{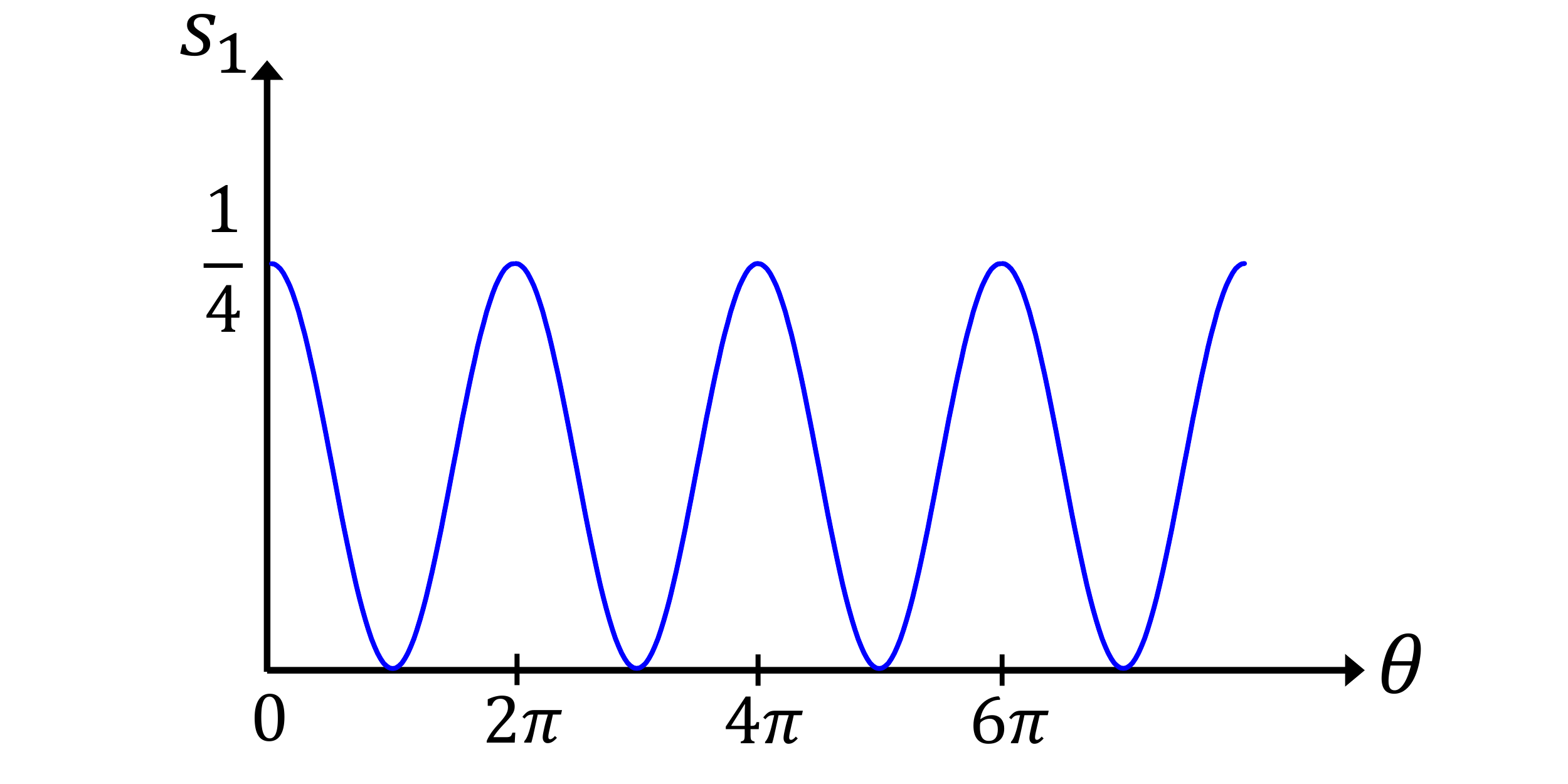}
\vspace{-0.2cm} 
\caption{\label{scheme} \small  ``Rabi flopping" in the simplex. Here we apply the Rabi rotation matrix of Eq.~(13) to the initial simplex vector with the initial condition of $x_0=1, y_0=1, x_1=0, y_1=0$ for the quantum wavefunction (and therefore the simplex amplitudes). As the rotation angle $\theta$ is varied, the first probability component of the simplex vector $s_1= \frac{1}{8} (1+x_0 \cos{\theta} - x_1 \sin{\theta})$ oscillates between the values of $0$ and $\frac{1}{4}$.  }
\end{center}
\vspace{-0.3cm} 
\end{figure}

\section{VII. Single-qubit phase gate}

The Quantum Mechanical single-qubit phase gate is accomplished by multiplying the state vector $| \psi \rangle$ with the following unitary matrix:
\begin{eqnarray}
\hat{U}_S = \left( \begin{array}{cc} 
1 & 0 \\
0 & \exp{(i \alpha)}  
\end{array} \right) = 
\left( \begin{array}{cc} 
1 & 0 \\
0 & \cos{\alpha} + i \sin{\alpha} 
\end{array} \right) 
\quad .
\end{eqnarray}

\noindent Note again that as required by unitarity, we have $\hat{U}_S^\dag \hat{U}_S=1$. The effect of the single-qubit phase gate on the quantum state is:
\begin{eqnarray}
\hat{U}_S | \psi \rangle & = &  \left( \begin{array}{cc} 
1 & 0 \\
0 & \cos{\alpha} + i \sin{\alpha} 
\end{array} \right) 
\left( \begin{array}{c} 
x_0 + i y_0 \\
x_1 + i y_1
\end{array} \right) \nonumber \\
&  =  & \left( \begin{array}{c} 
x_0 + i y_0 \\
\left( x_1 \cos{\alpha} - y_1 \sin{\alpha} \right) + i  \left( x_1 \sin{\alpha} + y_1 \cos{\alpha} \right) 
\end{array} \right)   \quad . 
\end{eqnarray}

\noindent By inspection, the required $ 8 \times 8$ matrix for this transformation is:
\begin{eqnarray}
\tilde{M}(\hat{U}_{S}) = \left( \begin{array}{cccccccc}
1 & 0 & 0 & 0 &  0 & 0 & 0 & 0 \\
0 & \cos\alpha & 0 & 0 & 0 & 0 & 0 & \sin\alpha \\ 
0 & 0 & 1 & 0 & 0 & 0 & 0 & 0 \\
0 & 0 & 0 & \cos\alpha & 0 & \sin\alpha & 0 & 0 \\
0 & 0 & 0 & 0 & 1 & 0 & 0 & 0 \\
0 & \sin\alpha & 0 & 0 & 0 & \cos{\alpha}  & 0  & 0 \\
0 & 0 & 0 & 0 & 0  & 0 & 1  & 0 \\
0 & 0 & 0 & \sin\alpha  & 0 & 0  & 0 &  \cos{\alpha}   \\
\end{array} \right) \quad . 
\end{eqnarray}

As we discussed above, with the matrix $\tilde{M}(\hat{U}_{S})$ given, the full transformation of the simplex vector which mimics single-qubit phase rotation is $ T_{\hat{U}_{S}}(\vec{s}) =  \frac{1}{8} \left[ \tilde{I} - \tilde{M}(\hat{U}_{S}) \right] \cdot \vec{1} + \tilde{M}(\hat{U}_{S}) \cdot \vec{s} $.

\section{VIII. Arbitrary evolution of the two-level quantum system}

Above, we argued that one can perform operations identical to the Hadamard gate and the single-qubit phase gate in the probability simplex. We also note that the transformations in the simplex rely on forming linear combinations of the vector, $\vec{p}$. As a result, a sequence of quantum gates can be mimicked by cascading the appropriate transformations in the simplex sequentially. More specifically, consider a quantum state that first evolves according to the unitary operator $\hat{U}_1$ followed by evolution with $\hat{U}_2$. We would find the corresponding transformation matrices $ \tilde{M}(\hat{U}_1)$ and $ \tilde{M}(\hat{U}_2)$, and map the corresponding vector as $\vec{p} \longrightarrow \tilde{M}(\hat{U}_2) \tilde{M}(\hat{U}_1) \cdot \vec{p}$. We, therefore, have $ \tilde{M}(\hat{U}_2 \hat{U}_1) = \tilde{M} (\hat{U}_2) \tilde{M} (\hat{U}_1 )$. We will provide a  more rigorous proof for this cascading theorem later in Section XII.

Because it is well-known that any arbitrary of evolution of a two-level quantum system can be accomplished using an appropriate combination of a Hadamard gate and a single-qubit phase gate \cite{nielsen}, we conclude that any arbitrary evolution can be achieved in the simplex. This result shows that both the initial state and the full evolution dynamics of a single qubit can be captured by using an 8-dimensional probabilistic simplex (or equivalently, using three classical probabilistic bits, since the simplex dimension of three classical probabilistic bits is 8). 

Below, in Section~XII, we will derive the differential equation for the time evolution of the simplex state $\vec{s}(t)$ in time $t$, analogous to the Schrodinger's equation for the quantum state $|\psi(t)\rangle$. The existence of a time evolution equation in the simplex also points to the same result; that is, full evolution dynamics of the quantum state can be captured in the simplex.

\section{IX. Specific implementation for a Hadamard gate}

A detailed discussion of how the above transformations can physically and practically be accomplished using classical probabilistic bits is beyond the scope of this work. It is likely that extensions of the recently demonstrated probabilistic bits (the so-called p-bits) will be feasible \cite{datta1,datta2}. In this section, we will discuss one specific approach to construct the Hadamard transformation in the simplex, which is summarized by the matrix of $\tilde{M}(\hat{U}_H)$ of Eq.~(10), with the full simplex transformation given by: $T_{\hat{U}_{H}}(\vec{s}) = \frac{1}{8} \left[ \tilde{I} - \tilde{M}(\hat{U}_H) \right] \cdot \vec{1} + \tilde{M}(\hat{U}_H) \cdot \vec{s}$. We discuss on the first row of this transformation; other rows can be implemented in a similar manner. Using this map, the first entry of the simplex vector will need to transform as:
\begin{eqnarray}  
\frac{1}{8} \left( 1+x_0 \right) \longrightarrow  \frac{1}{8} \left[ \frac{1+x_0}{\sqrt{2}} + \frac{1-x_1}{\sqrt{2}} + (1-\sqrt{2}) \right] \quad . 
\end{eqnarray}

We note that this transformation involves a constant offset added to the linear combination of the first and fourth entries in the initial simplex vector:  $\frac{1}{8} (1+x_0)$ and  $\frac{1}{8} (1-x_1)$, respectively.  Let's denote two random variables, $a$ and $b$, with event probabilities $p(a)$ and $p(b)$, respectively. We choose the event probabilities $p(a)$ and $p(b)$ to be exactly these first and fourth entries of the simplex vector:
\begin{eqnarray}
p(a) & = & \frac{1}{8} \left(1+x_0 \right) \quad , \nonumber \\
p(b) & = & \frac{1}{8} \left(1-x_1 \right) \quad .
\end{eqnarray} 

To implement the first row of the Hadamard gate, we need to construct a circuit, which will have two entries at its input, $p(a)$ and $p(b)$, and produce the following output:
\begin{eqnarray}
\frac{p(a)}{\sqrt{2}} + \frac{p(b)}{\sqrt{2}} + \frac{1}{8}  (1-\sqrt{2}) \quad . 
\end{eqnarray} 

One way to accomplish this would be to utilize the probability of the logical OR operation. Note that, the probability of $ p(a \lor b)$ is:
\begin{eqnarray}
p(a \lor b) = p(a) + p(b) - p(a \land b) \quad . 
\end{eqnarray}

\noindent Rewriting Eq.~(19) in a slightly different way:
\begin{eqnarray}
\frac{p(a)}{\sqrt{2}} + \frac{p(b)}{\sqrt{2}} + \frac{1}{8}(1-\sqrt{2}) = p(a) + p(b) + (\frac{1}{\sqrt{2}}-1) p(a) + (\frac{1}{\sqrt{2}}-1) p(b) +\frac{1}{8}(1-\sqrt{2}) \quad . 
\end{eqnarray} 

\noindent We see that if these two events are chosen to be such that their intersecting set satisfies $ p(a \land b) = (1-\frac{1}{\sqrt{2}}) p(a) + (1- \frac{1}{\sqrt{2}}) p(b) +\frac{1}{8}(\sqrt{2}-1)$, we have $p( a \lor b)$ giving the correct transformation. We note that the appearance of the logical OR and logical AND operations in above is further indication that we do not need to specifically focus on linear maps in the simplex (since logical operations are inherently nonlinear). 

\section{X. Multiple systems: the tensor product}

When we have more than one qubit, the Hilbert space  is given by the tensor product of the Hilbert space of the individual qubits; i.e., the wavefunctions will be of the form (for $N$ qubits):
\begin{eqnarray}
|\psi \rangle = |\psi_1 \rangle \otimes |\psi_2 \rangle  \otimes |\psi_3 \rangle  \cdots \otimes |\psi_N \rangle \quad .  
\end{eqnarray}

For probabilistic spaces, we combine multiple vectors in an identical way. This has been discussed and rigorously proven in Ref.~\cite{barrett1}; it is also implicit in the discussion of the mathematical structure of the probability theory by de Finetti \cite{definetti}. However, this feature of combining probability spaces is not widely known. It is usually assumed that the tensor product is a feature that is very special and specific to quantum mechanics. For probabilistic spaces, the combined vector in the simplex is given by:
\begin{eqnarray}
\vec{s} = \vec{s_1} \otimes \vec{s}_2 \otimes \vec{s}_3  \cdots \otimes \vec{s}_N \quad . 
\end{eqnarray}

\section{XI. Two-qubit controlled-not (CNOT) gate}

In this section, we discuss how to map the two-qubit controlled-not (CNOT) gate to the probabilistic simplex. When we have two qubits, the Hilbert space of the quantum system is spanned by states $ |00 \rangle$, $|01 \rangle$, 
$|10 \rangle$, and $|11 \rangle$.  We now have four complex coefficients that describe the state, i.e., our wavefunction is 
\begin{eqnarray}
| \psi \rangle = c_{00} |00 \rangle + c_{01} |01 \rangle + c_{10} |10 \rangle + c_{11} |11 \rangle \quad ,
\end{eqnarray}  

\noindent or equivalently in vector notation, we can express the wavefunction as:
\begin{eqnarray}
| \psi \rangle = 
\left( \begin{array}{c} 
c_{00} \\
c_{01} \\
c_{10} \\
c_{11}  
\end{array} \right)  \quad . 
\end{eqnarray}

\noindent The two-qubit CNOT gate maps $ |00 \rangle \rightarrow |00 \rangle $,  $ |01 \rangle \rightarrow |01 \rangle $, $ |10 \rangle \rightarrow |11 \rangle $, and  $ |11 \rangle \rightarrow |10 \rangle $. Here, the first bit is the control bit and the second bit is the target bit. In matrix notation, the gate is accomplished using the following unitary matrix:
\begin{eqnarray}
\hat{U}_{CNOT} = \left( \begin{array}{cccc}
1 & 0 & 0 & 0 \\
0 & 1 & 0 & 0 \\
0 & 0 & 0 & 1 \\
0 & 0 & 1 & 0 
\end{array} \right) \quad . 
\end{eqnarray}
\noindent Using our formalism, we map each qubit to a simplex of dimension 8: i.e., the corresponding simplex dimension for the multi-partite system, $\vec{s} = \vec{s}_1 \otimes \vec{s}_2$, is $8 \times 8 =64$. To make the notation analogous to the two-qubit Hilbert space, we write each simplex vector as:
\begin{eqnarray}
\vec{s}_ i  = \tilde{S}_{8 \times 8} \left( \begin{array}{c} 
\vec{0}_i \\
\vec{1}_i
\end{array} \right), \quad 
    \tilde{S}_{8\times 8}=\left( \begin{array}{cccccccc}
1 & 0 & 0 & 0 &  0 & 0 & 0 & 0 \\
0 & 0 & 0 & 0 & 1 & 0 & 0 & 0 \\ 
0 & 1 & 0 & 0 & 0 & 0 & 0 & 0 \\
0 & 0 & 0 & 0 & 0 & 1 & 0 & 0 \\
0 & 0 & 1 & 0 & 0 & 0 & 0 & 0 \\
0 & 0 & 0 & 0 & 0 & 0  & 1  & 0 \\
0 & 0 & 0 & 1 & 0  & 0 & 0  & 0 \\
0 & 0 & 0 & 0  & 0 & 0  & 0 &  1 \\
\end{array} \right) \quad .
\end{eqnarray}
\noindent Here, both $\vec{0}_i$ and $\vec{1}_i$ are column vectors with four entries: $\vec{0}_i=(1+x^{i}_0\,\,1-x^{i}_0\,\,1+y^{i}_0\,\,1-y^{i}_0)^{T}/8$ stores the real and imaginary parts of the complex coefficient for state $|0\rangle$ for the $i$'th qubit, while the vector $\vec{1}_i=(1+x^{i}_1\,\,1-x^{i}_1\,\,1+y^{i}_1\,\,1-y^{i}_1)^{T}/8$ stores the information for state $|1 \rangle$. Because of the unique way the mapping of Eq.~(2) is defined, we need the $8 \times 8$ ``shuffling" matrix $\tilde{S}_{8 \times 8}$ in Eq.~(27). The complete basis states for the two-qubit Hilbert space mapped to the simplex can similarly be written as:
\begin{eqnarray}
\vec{s} = \vec{s}_1 \otimes \vec{s}_2 =\tilde{S}_{64 \times 64}  \left( \begin{array}{c}
\vec{0}_1 \otimes \vec{0}_2 \\
\vec{0}_1 \otimes \vec{1}_2 \\
\vec{1}_1 \otimes \vec{0}_2 \\
 \vec{1}_1 \otimes \vec{1}_2 
\end{array} \right) \quad. 
\end{eqnarray}
\noindent Here, we note that each of these entries in the column of Eq.~(28) is of dimension $4 \times 4 =16$, and the total dimension is, $64$, as required. We have found that the new ``shuffling" matrix $\tilde{S}_{64\times 64}$ can be expressed as:
\begin{equation}
    \tilde{S}_{64\times 64}=(\tilde{S}_{8\times 8}\otimes\tilde{S}_{8\times 8})(\tilde{I}_{2\times 2}\otimes\tilde{S}_{8\times 8}\otimes\tilde{I}_{4\times 4}) \quad .
\end{equation}
Here, the quantities $\tilde{I}_{2 \times 2}$ and $\tilde{I}_{4 \times 4}$ are $2 \times 2$ and $4 \times 4$ identity matrices, respectively. The corresponding matrix that would implement the equivalent of CNOT gate would be a $64 \times 64$ matrix, acting on vectors with the form of above. In an analogous way to the CNOT gate of Eq.~(26), this $64 \times 64$ matrix will be:
\begin{eqnarray}
\tilde{M}_{CNOT} = \left( \begin{array}{cccc}
 \tilde{I}_{16 \times 16} & 0 & 0 & 0 \\
0 & \tilde{I}_{16 \times 16} & 0 & 0 \\
0 & 0 & 0 & \tilde{I}_{16 \times 16} \\
0 & 0 & \tilde{I}_{16 \times 16} & 0 
\end{array} \right) \quad . 
\end{eqnarray}

\noindent Here, the quantity $\tilde{I}_{16 \times 16}$ is the $16 \times 16$ identity matrix. With the matrix of Eq.~(30), we can perform an analogous operation on the simplex vector using the transformation, $\vec{s} \longrightarrow \tilde{S}_{64 \times 64} \tilde{M}_{CNOT} \tilde{S}_{64 \times 64}^{-1} \cdot \vec{s}$. We note that differing from the single-qubit discussion of above, the ``two-qubit" transformation of the simplex state is a linear transformation, and is a stochastic matrix.

We also note that, in general, with an initial separable simplex vector of the form $\vec{s} = \vec{s}_1 \otimes \vec{s}_2$, the transformation $\vec{s} \longrightarrow \tilde{S}_{64 \times 64} \tilde{M}_{CNOT} \tilde{S}_{64 \times 64}^{-1} \cdot \vec{s}$, would produce a simplex vector that can no longer be written as the tensor product of two individual simplex vectors (i.e., $\vec{s} \neq \vec{s}_1 \otimes \vec{s}_2$). Mathematically, this is identical to a CNOT gate producing entanglement in a two-qubit system.

In this section, we discussed the most straightforward way to map multiple qubits to the probability space. Each qubit is mapped to an 8-dimensional probabilistic vector, and the tensor product of qubits is mapped to a tensor product of these 8 dimensional simplex vectors. For example, as we detailed above, using this approach, two qubits are mapped to a probability space of dimension $8 \times 8=64$. However, it is clear that this manner of mapping multiple qubits to the simplex is not efficient. This is because, two qubits form a Hilbert space of dimension 4 (i.e., we have 4 complex coefficients), and using an extension of the mapping of Eq.~(2), we should need a $4 \times 4 =16$ dimensional simplex vector to store these 4 complex coefficients. Mathematically, this means that while the approach that we used here is intuitive and has a rigorous theoretical foundation as described in Ref.~\cite{barrett1}, it does not formally preserve the map $\varphi:H\rightarrow S$, i.e., $\varphi(\otimes_{i=1}^{N}\ket{\psi_i})\neq \otimes_{i=1}^{n}\vec{s}_i$.  We have found a more efficient way for combining simplex vectors that resembles the tensor product rule and simultaneously preserves the map $\varphi$. We will discuss this approach and its' specific application to the CNOT gate in Appendices A and B.  

\section{XII. Infinitesimal evolution: the Schrodinger's Equation for a single qubit}

To derive the Schrodinger's equation for infinitesimal evolution in the probability space, we will re-express both the mapping and transformations in a more compact form. We first rewrite the mapping $\varphi$ of Eq.~(2) explicitly in terms of the real and imaginary components of the wavefunction. Defining $\vec{x} \equiv \Re{\ket{\psi}}$ and $\vec{y} \equiv \Im{\ket{\psi}}$:  
\begin{equation}
\varphi(\ket{\psi})=\varphi(\vec{x}+i\vec{y})=\vec{s}=\frac{1}{8}\left(\begin{array}{c}
	1\\
	1\\
	1\\
	\vdots
\end{array}\right)+\frac{1}{8}\left(\begin{array}{r}
	\vec{x}\\
	-\vec{x}\\
	\vec{y}\\
	-\vec{y}
\end{array}\right)=\frac{1}{8}(\vec{1}+\vec{p}) \quad .  
\end{equation}
and in algebraic form, 
\begin{equation}
    \varphi(\ket{\psi})=\frac{1}{8}(\vec{1}+\vec{r}\otimes\Re{\ket{\psi}}+\vec{i}\otimes\Im{\ket{\psi}}),
\end{equation}
where, $\vec{r}=(1\,\,-1\,\,0\,\,0)^{T}$ and $\vec{i}=(0\,\,0\,\,1\,\,-1)^{T}$.

Given a general unitary matrix $\hat{U}$ acting on a quantum state vector $\ket{\psi}$, we note that the real and imaginary parts of the wavefunction will transform as: 
\begin{equation}
	\hat{U}\ket{\psi}=\left[\Re(\hat{U})+i\Im(\hat{U})\right](\vec{x}+i\vec{y})=\left[\Re(\hat{U})\vec{x}-\Im(\hat{U})\vec{y}\right]+i\left[ \Re(\hat{U})\vec{y}+\Im(\hat{U})\vec{x}\right] \quad . 
\end{equation}
\noindent Here, the quantities $\Re(\hat{U})$ and $\Im(\hat{U})$ are the real and imaginary components of the evolution operator $\hat{U}$, respectively. This implies that, under general unitary evolution, the real and imaginary parts of the wavefunction will evolve as:
\begin{eqnarray}
	\vec{x} & \longrightarrow & \left[\Re(\hat{U})\vec{x}-\Im(\hat{U})\vec{y} \right] \quad \nonumber \\
	\vec{y} & \longrightarrow & \left[ \Re(\hat{U})\vec{y}+\Im(\hat{U})\vec{x}\right] \quad . 
\end{eqnarray}
For the mapped vector $\vec{s}$ in the simplex, the above evolution of the real and imaginary parts of the wavefunction implies the following transformation of the vector $\vec{p}$:
\begin{equation}
	\vec{p}\longrightarrow \left(\begin{array}{c c c c}
		\Re(\hat{U})&O&O&\Im(\hat{U})\\
		O&\Re(\hat{U})&\Im(\hat{U})&O\\
		\Im(\hat{U})&O&\Re(\hat{U})&O\\
		O&\Im(\hat{U})&O&\Re(\hat{U})
	\end{array}\right)\left(\begin{array}{r}
		\vec{x}\\
		-\vec{x}\\
		\vec{y}\\
		-\vec{y}
	\end{array}\right)=\tilde{M}(\hat{U})\cdot\vec{p} \quad . 
\end{equation}
We also note that due to the structure of the $\vec{p}$ vector, the following two transformations are equivalent:
\begin{equation}
    \left(\begin{array}{c c c c}
		\Re(\hat{U})&O&O&\Im(\hat{U})\\
		O&\Re(\hat{U})&\Im(\hat{U})&O\\
		\Im(\hat{U})&O&\Re(\hat{U})&O\\
		O&\Im(\hat{U})&O&\Re(\hat{U})
	\end{array}\right)\left(\begin{array}{r}
		\vec{x}\\
		-\vec{x}\\
		\vec{y}\\
		-\vec{y}
	\end{array}\right)=\left(\begin{array}{c c c c}
		O&-\Re(\hat{U})&O&\Im(\hat{U})\\
		-\Re(\hat{U})&O&\Im(\hat{U})&O\\
		\Im(\hat{U})&O&O&-\Re(\hat{U})\\
		O&\Im(\hat{U})&-\Re(\hat{U})&O
	\end{array}\right)\left(\begin{array}{r}
		\vec{x}\\
		-\vec{x}\\
		\vec{y}\\
		-\vec{y}
	\end{array}\right) \quad . 
\end{equation}
As a result of the above equivalence, we consider both of the matrices to be equivalent definitions of the transformation $\tilde{M}(\hat{U})$ which is associated with a general evolution of the wavefunction by the operator $\hat{U}$:
\begin{equation}
	\tilde{M}(\hat{U})=\left(\begin{array}{c c c c}
		\Re(\hat{U})&O&O&\Im(\hat{U})\\
		O&\Re(\hat{U})&\Im(\hat{U})&O\\
		\Im(\hat{U})&O&\Re(\hat{U})&O\\
		O&\Im(\hat{U})&O&\Re(\hat{U})
	\end{array}\right)=\tilde{\mathcal{R}}\otimes\Re(\hat{U}) + \tilde{\mathcal{I}}\otimes\Im(\hat{U}) \quad ,
\end{equation}
or, 
\begin{equation}
	\tilde{M}(\hat{U})=\left(\begin{array}{c c c c}
		O&-\Re(\hat{U})&O&\Im(\hat{U})\\
		-\Re(\hat{U})&O&\Im(\hat{U})&O\\
		\Im(\hat{U})&O&O&-\Re(\hat{U})\\
		O&\Im(\hat{U})&-\Re(\hat{U})&O
	\end{array}\right)=-\tilde{\mathcal{I}}^2\otimes\Re(\hat{U}) + \tilde{\mathcal{I}}\otimes\Im(\hat{U}) \quad .
\end{equation}

\noindent Above we have used two $4 \times 4$ matrices $\tilde{\mathcal{R}}$ and $\tilde{\mathcal{I}}$ to simplify the notation:
\begin{equation}
	\tilde{\mathcal{R}}=\tilde{I}_{4\times4}=\left(\begin{array}{c c c c}
 		1&0&0&0\\
 		0&1&0&0\\
 		0&0&1&0\\
 		0&0&0&1
 	\end{array}\right), \quad \tilde{\mathcal{I}}=\left(\begin{array}{c c c c}
 	0&0&0&1\\
 	0&0&1&0\\
 	1&0&0&0\\
 	0&1&0&0
 \end{array}\right) \quad . 
\end{equation}

\noindent We note again that, given the transformation matrix associated with arbitrary evolution of the quantum state, $\tilde{M}(\hat{U})$, the full transformation of the simplex vector is given by $T_{\hat{U}}(\vec{s}) =  \frac{1}{8} \left[ \tilde{I} - \tilde{M}(\hat{U}) \right] \cdot \vec{1} + \tilde{M}(\hat{U}) \cdot \vec{s}$. 
We now rigorously prove the cascading theorem as follows:
\begin{align}
   &\nonumber \tilde{M}(\hat{U}_2)\tilde{M}(\hat{U}_1)=[\tilde{\mathcal{R}}\otimes\Re(\hat{U}_2)+\tilde{\mathcal{I}}\otimes\Im(\hat{U}_2)][\tilde{\mathcal{R}}\otimes\Re(\hat{U}_1)+\tilde{\mathcal{I}}\otimes\Im(\hat{U}_1)] \nonumber \\
   &=\tilde{\mathcal{R}}\otimes\Re(\hat{U}_2)\Re(\hat{U}_1)+\tilde{\mathcal{I}}\otimes(\Re(\hat{U}_2)\Im(\hat{U}_1)+\Im(\hat{U}_2)\Re(\hat{U}_1))+\tilde{\mathcal{I}}^2\otimes\Im(\hat{U}_2)\Im(\hat{U}_1) \nonumber \\
   &\equiv\tilde{\mathcal{R}}\otimes(\Re(\hat{U}_2)\Re(\hat{U}_1)-\Im(\hat{U}_2)\Im(\hat{U}_1))+\tilde{\mathcal{I}}\otimes(\Re(\hat{U}_2)\Im(\hat{U}_1)+\Im(\hat{U}_2)\Re(\hat{U}_1)) \nonumber \\
   &=\tilde{\mathcal{R}}\otimes\Re(\hat{U}_2\hat{U}_1)+\tilde{\mathcal{I}}\otimes\Im(\hat{U}_2\hat{U}_1)=\tilde{M}(\hat{U}_2\hat{U}_1)\quad .
\end{align}
In this derivation, we have used the equivalence established above in Eqs.~(37) and (38).

In order to derive the analogue of Schrodinger's equation in the simplex, we follow the above formalism for infinitesimal evolution of the quantum state $\ket{\psi(t)}$, under the presence of some general Hamiltonian $\hat{H}(t)$. The time evolution of the wavefunction, which is determined by the Schrodinger's equation, is:
\begin{equation}
	i\hbar\frac{d}{dt}\ket{\psi(t)}=\hat{H}(t)\ket{\psi(t)} \quad .
\end{equation}
The transformation of the state from $\ket{\psi(t)}$ to $\ket{\psi(t+\delta t)}$ is due to the action of the differential unitary operator $\delta \hat{U}=\exp[-i\hat{H}(t)/\hbar\,\delta t]$. Correspondingly the differential transformation $\tilde{M}(\delta\hat{U})$ must act on the vector $\vec{p}(t)$, to produce a vector $\vec{p}(t+\delta t)$:
\begin{equation}
	\vec{p}(t+\delta t)=\tilde{M}(\delta \hat{U})\cdot\vec{p}(t)=\tilde{M}[\hat{I}-i\hat{H}(t)/\hbar\, \delta t]\cdot\vec{p}(t)=\vec{p}(t)-\frac{\delta t}{\hbar}\tilde{M}(i\hat{H}(t))\cdot\vec{p}(t) \quad . 
\end{equation}
We then evaluate $\tilde{M}(i\hat{H}(t))$ in Eq.~(42) using Eq.~(35) and take the $\delta t \rightarrow 0$ limit to obtain the following differential equation for $\vec{p}(t)$: 
\begin{equation}
	\hbar\frac{d\vec{p}(t)}{dt}=[\tilde{\mathcal{R}}\otimes\Im(\hat{H}(t))-\tilde{\mathcal{I}}\otimes\Re(\hat{H}(t))] \cdot \vec{p}(t) \equiv \tilde{H}_{eff}(t)\cdot\vec{p}(t) \quad . 
\end{equation}
\noindent Here, we have defined $\tilde{H}_{eff}(t)=\tilde{\mathcal{R}}\otimes\Im(\hat{H}(t))-\tilde{\mathcal{I}}\otimes\Re(\hat{H}(t))$, which can be viewed as the effective ``Hamiltonian" operator in the simplex. We next use Eq.~(43), to find the differential equation for the full simplex vector, $\vec{s}(t)$. Using  $\vec{s}(t)=\frac{1}{8} \left[ \vec{1}+\vec{p}(t) \right]$, we have  $\dot{\vec{s}}=\frac{1}{8}\dot{\vec{p}}$ for the time derivative of the full simplex vector. Using this time derivative in Eq.~(43), we finally arrive at:
\begin{eqnarray}
\hbar\frac{d\vec{s}(t)}{dt} =\tilde{H}_{eff}(t)\cdot(\vec{s}-\frac{1}{8}\vec{1}) \quad .
\end{eqnarray}

\section{XIII. Schrodinger's Equation for a general quantum state of dimension K}

Equation (44) is valid for the explicit mapping of a single qubit to an 8 dimensional probability simplex of Eq.~(2). We note that it is straightforward to generalize this result. For a Hilbert space of dimension $K$, we would have $K$ complex coefficients to describe an any wavefunction that lies in this space. We can then map the real and imaginary parts of these coefficients to a probability simplex of dimension $4K$. 

Specifically, for a wavefunction of dimension $K$, we can define its' real and imaginary parts in exactly the same manner $\vec{x}_K \equiv \Re{\ket{\psi}}$ and $\vec{y}_K \equiv \Im{\ket{\psi}}$. The map $\varphi(\ket{\psi})$ is then given by a similar expression as Eq.~(31) of above:
\begin{equation}
\varphi(\ket{\psi})=\varphi(\vec{x}_K+i\vec{y}_K)=\vec{s}=\frac{1}{4K}\left(\begin{array}{c}
	1\\
	1\\
	1\\
	\vdots
\end{array}\right)+\frac{1}{4K}\left(\begin{array}{r}
	\vec{x}_K\\
	-\vec{x}_K\\
	\vec{y}_K\\
	-\vec{y}_K
\end{array}\right)=\frac{1}{4K}(\vec{1}_{K}+\vec{p}) \quad .  
\end{equation}

Following the identical steps outlined above in Section~XII, we can then derive the time evolution equation for the $4K$ dimensional simplex state:
\begin{eqnarray}
\hbar\frac{d\vec{s}(t)}{dt} =\tilde{H}_{eff}(t)\cdot(\vec{s}-\frac{1}{4K}\vec{1}_{K}) \quad .
\end{eqnarray}

We note that the time evolution of Eq.~(46) preserves the probabilistic norm of the simplex vector, that is $\sum_{i}s_i(t)=1$ at all times. This can be proven using:
\begin{align}
	&\nonumber \frac{d \vec{s}}{dt}=\frac{1}{4K}\frac{d \vec{p}}{dt} \implies \vec{1}_{K}^{T}\cdot\frac{d \vec{s}}{dt}=\frac{1}{4K}\vec{1}_{K}^{T}\cdot\frac{d \vec{p}}{dt}=\frac{1}{4K\hbar}\vec{1}_{K}^{T}\cdot\tilde{H}_{eff}\cdot\vec{p}=0 \quad , \\
	&\therefore\frac{d}{dt}(\sum_{i}s_i(t))=0\implies \sum_{i}s_i(t)=\sum_{i}s_i(0)=1 \quad .
\end{align}

\section{XIV. Relation to prior work}

Since the inception of quantum mechanics, the precise meaning of the wavefunction $| \psi \rangle $ has been a source of constant debate. Does the wavefunction correspond to something real (the ontological interpretation) \cite{montina,rudolph2,rudolph3}, or is it simply a tool to summarize an observer's incomplete information about a system (the epistemological interpretation) \cite{spekkens1}. This work firmly supports the epistemological interpretation of the wavefunction. We have shown that the initial state and the dynamics of a single qubit can be completely captured using an 8-dimensional probabilistic simplex. It is the thesis of this work the traditional mathematical formulation of quantum theory, i.e., complex vectors that live in a Hilbert space, is not needed. The problem can instead be formulated as a real vector in a probability space and traditional features of quantum theory, such as interference of complex amplitudes, can be captured in this space. As also discussed by other authors \cite{barrett1}, the identical nature of the tensor product for combining separate systems, also strongly favors the epistemological interpretation. 

Within this context, perhaps the closest relation to our work is the ``toy" theory of Spekkens, which captures some features of a quantum two-level system using four epistemological states, based on the knowledge balance principle \cite{spekkens1}. Ontological, hidden variable theories of quantum mechanics has also been widely discussed \cite{rudolph1}. Several prominent examples of these are due to Bell \cite{bell2}, Beltrametti-Bugajski \cite{bugajski}, Kochen-Specker \cite{kochen}, Aaronson \cite{aaronson}, and Aerts \cite{aerts}. Of particular importance to this work is Aaronson's model \cite{aaronson}, which discusses representing the quantum state as a vector of probabilities, and mapping this vector to another set of probabilities using an appropriate matrix. However, as we mentioned above, when only represented as a vector of projected probabilities, such a matrix inevitably depends on the initial state of the wavefunction. We emphasize again that this is very different from our approach and our results. The analogs of the quantum gates that we have discussed in the probability space, such as the Hadamard gate of Eq.~(10) and the single-qubit phase gate of Eq.~(16), involve constant translations and constant matrices (i.e., the information about the initial state, $x_0, y_0, x_1, y_1$ does not appear in these matrices). We believe this is crucial for the approach presented here. Consider a quantum system that evolves through any number of gates. One can construct probabilistic circuits that would perform exact analogs of these gates in the probability simplex, and exactly track the quantum evolution in the simplex. Once the circuits are fixed, if the initial condition for the quantum state changes, one only needs to correspondingly modify the initial simplex vector.

We also note that our work has connection to the work of Fuchs and colleagues, who has argued for a completely subjective interpretation of the quantum wavefunction \cite{fuchs1,fuchs2,fuchs3}. In this approach, the key ingredient is a set of projection operators, which are called SICs (abbreviation for Symmetic Informationally Complete). These operators have a number of fine-tuned properties. Of particular importance, when one expresses the density matrix operator for a quantum system as a summation of SICs, the expansion coefficients explicitly have the measurement outcome probabilities. As a result, these operators allow for a smooth relation between the measurement outcome probabilities and the density matrix. There are no other operator classes that allow such a simple and explicit connection between measurement probabilities and the density matrix for a quantum system. 

We note that here are several key differences between our work and the approach of SIC operators. (1) The SIC approach focuses on the measurement outcomes of a quantum mechanical system. As we mentioned above, our manuscript focuses on capturing the initial state and the dynamics of a quantum system. (2) To our knowledge, our approach is unique, in the sense that it allows storing both the amplitude and phase information of the complex coefficients of a quantum system. This is critical for capturing the single-qubit and two-qubit gates in the probability simplex, as well as finding the equivalent of the Schrodinger’s equation in the probability space.  

Our work also has an important connection to the large body of literature that have studied mapping of quantum states to quasiprobability distributions. The most famous of these is the Wigner function which is in the continuous phase space. Recent work has focused on negativity of these quasiprobability distributions, their studies in discrete phase spaces, as well as their relationship to quantum contextuality \cite{bartlett1,bartlett2,bartlett3,raussendorf1,raussendorf2,eisert,zhu,wootters}. It is known that such quasiprobability distributions can have negative values, and the amount of negativity is typically viewed as the ``quantumness" of the system.  The key difference of our work is that we have discussed that if one considers appropriate maps,  then  mapping to a classical probabilistic space is possible (i.e., negative values are not needed). 

We also note that in this work, we have focused on the initial state and the time evolution of quantum systems and how we can capture these in the probability space: i.e., we have not considered the measurement aspect of quantum systems. More specifically, the focus of the current work is on capturing the initial quantum states and the gates, largely within the context of an $N$-qubit quantum computer. One key reason for this focus is that when one analyzes a quantum algorithm in detail (for example, Shor factoring algorithm \cite{nielsen}), the key advantage is in the interference of complex amplitudes in an exponentially large Hilbert space (i.e., the key advantage is not in the measurement; the measurement in the end is more like an afterthought). We also have not gone into a detailed discussion of entanglement \cite{bell1}, nonlocality \cite{barrett2}, or contextuality \cite{spekkens2}. A rigorous discussion of these important concepts is beyond the scope of this work. 

\section{XV. Conclusions and future directions}

In conclusion, we have discussed mapping of the initial state and time evolution of a qubit to the probability simplex. We have discussed analogs of the single-qubit Hadamard and phase gates, as well as two-qubit CNOT gate in the probabilistic simplex. We have also discussed how multi-partite quantum systems can be mapped, as well as the analog of the Schrodinger's equation which dictates the time evolution of the simplex vector. 

It is well known that the quantum gates that we have discussed here (single-qubit Hadamard gate, single-qubit phase rotation, and the two-qubit CNOT gate) form a universal gate set \cite{nielsen}. That is, arbitrary evolution of an $N$ qubit quantum computer can be approximated with these gates. Due to the similar nature of the mathematical formalism, perhaps a similar result can be expected for $N$ simplex vectors: but this is an open question. 

We believe that the approach presented here may provide a unique way to simulate quantum systems that are more efficient than currently possible. For example, for an $N$-qubit quantum computer, the initial state at the beginning of a computation would be an unentangled product state of the individual qubits. We would then map each qubit to an 8-dimensional probabilistic vector, $\vec{s}$. This could be any 8-dimensional probabilistic system, whose probabilities (individual entries in the vector) have appropriate values so that the initial state of each qubit is accurately mapped. The tensor product of $N$ qubits would then map to a tensor product of $N$ 8-dimensional probabilistic vectors. The evolution of the quantum computer would proceed as a sequence of single qubit and two-qubit gates. The simulation that we envision then is to implement the analogs of these gates in the probabilistic system, using the appropriate transformations in the simplex that we have described above. It is an open question how much $N$ simplex vectors (forming a probability space of dimension $8^N$) can efficiently simulate an $N$ qubit quantum computer (a Hilbert space of dimension $2^N$). 

Within this context, an exciting immediate experimental direction is to experimentally demonstrate the simplex transformations for a single qubit that we have discussed. It may be possible to extend the recent experimental work of Datta and colleagues on probabilistic bits (p-bits) \cite{datta1,datta2}. One key goal would be to observe the ``Rabi flopping" behavior of Fig.~2 in the simplex using an appropriate circuit acting on 3~p-bits. 

We think it is also possible that progress along the above posed questions will help clarify the quantum/classical boundary \cite{niklas,mucciolo,pittenger}, as well as the quantum measurement problem \cite{zurek1,zurek2}.

\section{XVI. Acknowledgements}

We would like to thank Ben Lemberger and Volkan Rodoplu for many helpful discussions. D. D. Yavuz would also would like to thank Bin Yan for an early discussion on the subject. This work was supported by the National Science Foundation (NSF) Grant No. 2016136 for the QLCI center Hybrid Quantum Architectures and Networks (HQAN), and also by the University of Wisconsin-Madison, through the Vilas Associates award. 

\newpage

\section{Appendix A. Multiple systems: The tensor product revisited}

Above in Section~X, we discussed the most straightforward way to map multiple qubits to the probability space. Each qubit is mapped to an 8-dimensional probabilistic vector, and the tensor product of qubits is mapped to a tensor product of these 8 dimensional simplex vectors. For example, as we detailed above, using this approach, two qubits are mapped to a probability space of dimension $8 \times 8=64$. However, it is clear that this mapping is not efficient. This is because, two qubits form a Hilbert space of dimension 4 (i.e., we have 4 complex coefficients), and using an extension of the mapping of Eq.~(2), we should need a $4 \times 4 =16$ dimensional simplex vector to store these 4 complex coefficients. Mathematically, this means while the approach that we used in Section~X is intuitive, it does not formally preserve the map $\varphi:H\rightarrow S$, i.e., $\varphi(\otimes_{i=1}^{N}\ket{\psi_i})\neq \otimes_{i=1}^{n}\vec{s}_i$.

In this section, we discuss an approach to combine multiple simplex vectors that preserves the map and also closely resembles the tensor product structure.  
For this purpose, we first look at the combination scheme of the real and imaginary parts of the quantum states for the two qubits, $\ket{\psi_1}$ and $\ket{\psi_2}$, since the definition of map $\varphi:H\rightarrow S$ in Eq.~(32) explicitly includes the real and imaginary components. The real and imaginary parts for the wavefunction of the tensor product is: 
\begin{equation}
	\ket{\psi_1}\otimes\ket{\psi_2}=(\vec{x}_1+i\vec{y}_1)\otimes(\vec{x}_2+i\vec{y}_2)=(\vec{x}_1\otimes\vec{x}_2-\vec{y}_1\otimes\vec{y_2})+i(\vec{x_1}\otimes\vec{y}_2+\vec{y}_1\otimes\vec{x}_2) \quad . 
\end{equation}
Here, we have defined the real and imaginary parts of the wavefunctions for each qubit: $\vec{x}_1 \equiv \Re{\ket{\psi_1}}$, $\vec{y}_1 \equiv \Im{\ket{\psi_1}}$, $\vec{x}_2 \equiv \Re{\ket{\psi_2}}$, and $\vec{y}_2 \equiv \Re{\ket{\psi_2}}$. Using an extension of the map of Eq.~(32), this then implies:
\begin{align}
	\nonumber\varphi(\ket{\psi_1}\otimes\ket{\psi_2})=\frac{1}{16}(\vec{1}_{16}&+\vec{r}\otimes[\Re{\ket{\psi_1}}\otimes\Re{\ket{\psi_2}}-\Im{\ket{\psi_1}}\otimes\Im{\ket{\psi_2}}]\\
	&+\vec{i}\otimes[\Re{\ket{\psi_1}}\otimes\Im{\ket{\psi_2}}+\Im{\ket{\psi_1}}\otimes\Re{\ket{\psi_2}}]) \quad . 
\end{align}
We note that the above equation is precisely equal to $\frac{1}{16}(\vec{1}_{16}+\vec{r}\otimes\Re{\ket{\psi_1}\otimes\ket{\psi_2}}+\vec{i}\otimes\Im{\ket{\psi_1}\otimes\ket{\psi_2}})$. Hence we introduce a closed operation which we refer to as the ``box product",  $\boxtimes$, over any two $4m$ ($m\ge2,m\in\mathbb{N}$) dimensional vectors that have the form of a typical $\vec{p}$ to preserve the map while combining systems, 
\begin{equation}
	\vec{p}_1\boxtimes\vec{p}_2=\left(\begin{array}{r}
		\vec{x}_1\\
		-\vec{x}_1\\
		\vec{y}_1\\
		-\vec{y}_1
	\end{array}\right)\boxtimes\left(\begin{array}{r}
	\vec{x}_2\\
	-\vec{x}_2\\
	\vec{y}_2\\
	-\vec{y}_2
	\end{array}\right) \equiv \left(\begin{array}{r}
	(\vec{x}_1\otimes\vec{x}_2-\vec{y}_1\otimes\vec{y_2})\\
	-(\vec{x}_1\otimes\vec{x}_2-\vec{y}_1\otimes\vec{y_2})\\
	(\vec{x_1}\otimes\vec{y}_2+\vec{y}_1\otimes\vec{x}_2)\\
	-(\vec{x_1}\otimes\vec{y}_2+\vec{y}_1\otimes\vec{x}_2)
\end{array}\right) \quad . 
\end{equation}
We note that this operation is closed because the resultant vector, $\vec{p}_1 \boxtimes \vec{p}_2$ follows the typical form of the $\vec{p}$ vector. We also note that the above defined box product operation remains closed even if it is extended to the matrices of the usual form $\tilde{M}(\hat{U})$ for any unitary operator $\hat{U}$. that acts on the composite quantum system, $\hat{U} = \hat{U}_1 \otimes \hat{U}_2$. To prove this we start with: 
\begin{equation}
	\tilde{M}(\hat{U}_1\otimes\hat{U}_2)(\vec{p}_1\boxtimes\vec{p}_2)=\tilde{M}(\hat{U}_1)\boxtimes\tilde{M}(\hat{U}_2)(\vec{p}_1\boxtimes\vec{p}_2)=\tilde{M}(\hat{U}_1)\vec{p}_1\boxtimes\tilde{M}(\hat{U}_2)\vec{p}_2 \quad . 
\end{equation} 
We then expand the rightmost expression and equate it  term by term with the middle term to obtain the following form for $\tilde{M}(\hat{U}_1)\boxtimes\tilde{M}(\hat{U}_2)$:
\begin{align}
	\nonumber \tilde{M}(\hat{U}_1)\boxtimes\tilde{M}(\hat{U}_2)=\tilde{\mathcal{R}}\otimes[\Re(\hat{U}_1)\otimes\Re(\hat{U}_2)-&\Im(\hat{U}_1)\otimes\Im(\hat{U}_2)]+\\
	&\tilde{\mathcal{I}}\otimes[\Re(\hat{U}_1)\otimes\Im(\hat{U}_2)+\Im(\hat{U}_1)\otimes\Re(\hat{U}_2)] \quad . 
\end{align}
The expression in Eq.~(52) is exactly equal to  $\tilde{M}(\hat{U}_1\otimes\hat{U}_2)$. This is because  $\Re(\hat{U}_1\otimes\hat{U}_2)\neq \Re(\hat{U}_1)\otimes\Re(\hat{U}_2)$, but instead  $\Re(\hat{U}_1\otimes\hat{U}_2)=\Re(\hat{U}_1)\otimes\Re(\hat{U}_2)-\Im(\hat{U}_1)\otimes\Im(\hat{U}_2)$. Similarly, we also have $\Im(\hat{U}_1\otimes\hat{U}_2)=\Re(\hat{U}_1)\otimes\Im(\hat{U}_2)+\Im(\hat{U}_1)\otimes\Re(\hat{U}_2)$.
Hence, in the above defined probabilistic simplex space, $S$, the combined vector in the simplex is given by:
\begin{eqnarray}
\vec{s} = \frac{1}{2^{N+2}}(\vec{1}_{2^{N+2}}+((((\vec{p}_1 \boxtimes \vec{p}_2) \boxtimes \vec{p}_3 ) \cdots) \boxtimes \vec{p}_N)) \quad . 
\end{eqnarray}
One important distinctive property of the box product operation  versus the tensor product operation is that it is distributive over addition but not associative (just like the universal logical operations NAND and NOR). 

\section{Appendix B. Two-qubit controlled-not (CNOT) gate revisited}

In this section, we will discuss how to map the two-qubit controlled-not gates to the probability simplex, using the techniques introduced in the previous section, in Appendix~A. As we discussed above, when we have two qubits, the Hilbert space of the quantum system is of dimension four and it is spanned by states $ |00 \rangle$, $|01 \rangle$, $|10 \rangle$, and $|11 \rangle$. We can, therefore, represent the wavefunction as $\ket{\psi}=\vec{x}+i\vec{y}$, where $\vec{x}$ and $\vec{y}$ are each of vectors of dimension $4$. In the box product formalism, the dimension of the simplex space would be $4\times4=16$ and in that space the matrix of transformation would be $\tilde{M}(\hat{U}_{CNOT})=\tilde{R}\otimes\hat{U}_{CNOT}$, since $\hat{U}_{CNOT}$ is real,
\begin{equation}
	\tilde{M}(\hat{U}_{CNOT})=\left( \begin{array}{cccc}
		\hat{U}_{CNOT} & O & O & O\\
		O & \hat{U}_{CNOT} & O & O \\
		O & O & \hat{U}_{CNOT} & O \\
		O & O & O & \hat{U}_{CNOT} 
	\end{array} \right) \quad . 
\end{equation}

\noindent As our next example, we consider a general conditional unitary gate which can be written as: $\hat{C}_{U}=\hat{P}_{0}\otimes\hat{I}+\hat{P}_1\otimes\hat{U}$. The two-qubit transformation matrix for this gate, $\tilde{M}(\hat{C}_U)$ is given by:
\begin{align}
	&\nonumber\tilde{M}(\hat{C}_{U})=\tilde{M}(\hat{P}_{0}\otimes\hat{I}+\hat{P}_1\otimes\hat{U})=\tilde{M}(\hat{P}_{0}\otimes\hat{I})+\tilde{M}(\hat{P}_{1}\otimes\hat{U})\\
	\nonumber\Rightarrow&\tilde{M}(\hat{C}_{U})=\tilde{\mathcal{R}}\otimes[\hat{P}_0\otimes\hat{I}]+\tilde{\mathcal{R}}\otimes[\hat{P}_1\otimes\Re(\hat{U})]+\tilde{\mathcal{I}}\otimes[\hat{P}_1\otimes\Im(\hat{U})]\\
	\Rightarrow&\tilde{M}(\hat{C}_{U})=\tilde{\mathcal{R}}\otimes\hat{C}_{\Re(U)}+\tilde{\mathcal{I}}\otimes\hat{P}_1\otimes\Im(\hat{U}) \quad . 
\end{align}
In matrix form this can be represented as:
\begin{equation}
	\tilde{M}(\hat{C}_{U})=\left( \begin{array}{cccc}
		\hat{C}_{\Re(U)} & O & O & \hat{P}_1\otimes\Im(\hat{U})\\
		O & \hat{C}_{\Re(U)} & \hat{P}_1\otimes\Im(\hat{U}) & O \\
		\hat{P}_1\otimes\Im(\hat{U}) & O & \hat{C}_{\Re(U)} & O \\
		O & \hat{P}_1\otimes\Im(\hat{U}) & O & \hat{C}_{\Re(U)} 
	\end{array} \right) \quad . 
\end{equation}

\end{document}